# Validity of altmetrics data for measuring societal impact: A study using data from Altmetric and F1000Prime

Lutz Bornmann

Division for Science and Innovation Studies

Administrative Headquarters of the Max Planck Society

Hofgartenstr. 8,

80539 Munich, Germany.

E-mail: bornmann@gv.mpg.de

**Abstract**

Can altmetric data be validly used for the measurement of societal impact? The current study seeks to answer this question with a comprehensive dataset (about 100,000 records) from very disparate sources (F1000, Altmetric, and an in-house database based on Web of Science). In the F1000 peer review system, experts attach particular tags to scientific papers which indicate whether a paper could be of interest for science or rather for other segments of society. The results show that papers with the tag "good for teaching" do achieve higher altmetric counts than papers without this tag – if the quality of the papers is controlled. At the same time, a higher citation count is shown especially by papers with a tag that is specifically scientifically oriented ("new finding"). The findings indicate that papers tailored for a readership outside the area of research should lead to societal impact.

If altmetric data is to be used for the measurement of societal impact, the question arises of its normalization. In bibliometrics, citations are normalized for the papers' subject area and publication year. This study has taken a second analytic step involving a possible normalization of altmetric data. As the results show there are particular scientific topics which are of especial interest for a wide audience. Since these more or less interesting topics are not completely reflected in Thomson Reuters' journal sets, a normalization of altmetric data should not be based on the level of subject categories, but on the level of topics.

# 1   Introduction

In science policy it was assumed into the 1990s that society can benefit most from a science which pursues research at a high level. Correspondingly, indicators were (and are) used in scientometrics, such as citation counts, which measure the impact of research on science itself. Since the 1990s a trend can be observed in science policy no longer to assume that society benefits from a science pursued at a high level (Bornmann, 2012, 2013). It is now expected that the benefit for society be demonstrated. Thus, for example, organizations which support research (such as, for example, the US National Science Foundation) now expect that supported projects lead to an outcome which is of interest not solely to science. For these organizations the consequence for the peer review procedure is that not only the possible scientific yield of the project has to be assessed, but also the returns for other sections of society.

These days, scientific work is not assessed solely on the basis of the peer review procedure, but also with indicators. A good example of these quantitative assessments is university ranking (Hazelkorn, 2011). The most important indicators in this connection (not only with university ranking) are bibliometric indicators based on publications and their citations (Vinkler, 2010). The impact of research is generally measured with citations. Since the impact of one publication on another publication is measured here, citations measure the impact of research on research itself. Citations allow a determination as to whether research (for example in institutions or countries) is being pursued at the highest level on average or not. But citations cannot be used to measure the impact of research on other sections of society. This is why scientometrics has taken up the wish in science policy to measure the impact of research beyond the confines of science, and is seeking new possibilities for impact measurement (Bornmann, 2014). With societal impact assessments the (1) social, (2) cultural, (3) environmental and (4) economic returns (impact and effects) from results (research output) or products (research outcome) of publicly funded research are measured (Bornmann, 2013). Currently the most favored procedure for measuring societal impact involves case studies, which, however, are seen as too time-consuming and therefore less practicable.

An attractive possibility for measuring societal impact is seen in altmetrics (short for alternative metrics) (Mohammadi & Thelwall, 2014). "Altmetrics refers to data sources, tools, and metrics (other than citations) that provide potentially relevant information on the impact of scientific outputs (e.g., the number of times a publication has been tweeted, shared on Facebook, or read in Mendeley). Altmetrics opens the door to a broader interpretation of the concept of impact and to more diverse forms of impact analysis" (Waltman & Costas, 2014, p. 433). An overview of various altmetrics may be obtained from Priem and Hemminger (2010). Twitter (www.twitter.com), for example, is the best known microblogging application. This application allows the user to post short messages (tweets) of up to 140 characters. "These tweets can be categorized, shared, sent directly to other users and linked to websites or scientific papers … Currently there are more than 200 million active Twitter users who post over 400 million tweets per day" (Darling, Shiffman, Côté, & Drew, 2013). Priem and Costello (2010) define tweets as Twitter citations if they contain a direct or indirect link to a peer-reviewed scholarly article. These Twitter citations can be counted and assessed as an alternative metric for papers.

There are already a number of studies concerning altmetrics. An overview of these studies can be found in Bar-Ilan, Shema, and Thelwall (2014), Haustein (2014), and Priem (2014). Many of these studies have measured the correlation between citations and altmetrics. Since the correlations were often at a moderate level, the results are difficult to interpret: Both metrics seem to measure something similar but not identical. The studies published so far cannot yet provide a satisfactory answer to the question whether altmetrics is appropriate for the measurement of societal impact or not. That is the reason for this investigation of the question.

In January 2002, a new type of peer-review system has been launched, in which about 5000 Faculty members are asked "to identify, evaluate and comment on the most interesting papers they read for themselves each month – regardless of the journal in which they appear" (Wets, Weedon, & Velterop, 2003, p. 251). What is known as the Faculty of 1000 (F1000) peer review system is accordingly not an ex-ante assessment of manuscripts provided for publication in a journal, but an ex-post assessment of papers which have already been published in journals. The Faculty members also attach tags to the papers indicating their relevance for science (e.g. "new finding"), but which can also serve other purposes. One example of the tags which the members can attach is "good for teaching". Papers can be marked in this way if they represent a key paper in a field, are well written, provide a good overview of a topic, and/or are well suited as literature for students. Papers marked with this tag can be expected to have an impact beyond science itself (that means societal impact), unlike papers without this tag. If altmetrics indicate a greater impact for papers with this tag than those without, this would suggest that altmetrics measure societal impact.

This study is essentially based on a dataset with papers (and their evaluations and tags from Faculty members) extracted from F1000 (see also Mohammadi & Thelwall, 2013). This dataset was extended with further data – bibliometric (e.g. citation counts) and altmetric (e.g. Twitter counts). There follows in the next sections a comparison of altmetric counts with citation counts, to investigate the differences between the two metrics in relation to tags and recommendations.

## 2 Methods

### 2.1 Peer ratings provided by F1000

F1000 is a post-publication peer review system of the biomedical literature (papers from medical and biological journals). This service is part of the Science Navigation Group, a group of independent companies that publish and develop information services for the professional biomedical community and the consumer market. F1000 Biology was launched in 2002 and F1000 Medicine in 2006. The two services were merged in 2009 today constitute the F1000 database. Papers for F1000 are selected by a peer-nominated global "Faculty" of leading scientists and clinicians who then rate them and explain their importance (F1000, 2012). This means that only a restricted set of papers from the medical and biological journals covered is reviewed, and most of the papers are actually not (Kreiman & Maunsell, 2011; Wouters & Costas, 2012).

The Faculty nowadays numbers more than 5,000 experts worldwide, assisted by 5,000 associates, which are organized into more than 40 subjects (which are further subdivided into over 300 sections). On average, 1,500 new recommendations are contributed by the Faculty each month (F1000, 2012). Faculty members can choose and evaluate any paper that interests them; however, "the great majority pick papers published within the past month, including advance online papers, meaning that users can be made aware of important papers rapidly" (Wets, et al., 2003, p. 254). Although many papers published in popular and high-profile journals (e.g. *Nature*, *New England Journal of Medicine*, *Science*) are evaluated, 85% of the papers selected come from specialized or less well-known journals (Wouters & Costas, 2012). "Less than 18 months since Faculty of 1000 was launched, the reaction from scientists has been such that two-thirds of top institutions worldwide already subscribe, and it was the recipient of the Association of Learned and Professional Society Publishers (ALPSP) award for Publishing Innovation in 2002 (http://www.alpsp.org/about.htm)" (Wets, et al., 2003, p. 249). The F1000 data base is regarded as a significant aid for scientists seeking the most relevant papers in their subject area: "The aim of Faculty of 1000 is not to provide an evaluation for all papers, as this would simply exacerbate the 'noise', but to take advantage of electronic developments to create the optimal human filter for effectively reducing the noise," (Wets, et al., 2003, p. 253).

The papers selected for F1000 are rated by the members as "Good," "Very good" or "Exceptional" which is equivalent to scores of 1, 2, or 3, respectively. In many cases a paper is assessed not just by one member but by several. The FFa (F1000 Article Factor), given as a total score in the F1000 database, is calculated from the different recommendations for a publication. Besides making recommendations, Faculty members also tag publications with classifications, as for example (see http://f1000.com/prime/my/about/evaluating):

- Clinical Trial (non-RCT): investigates the effects of an intervention (but neither randomized nor controlled) in human subjects.

- Confirmation: the findings of the article validate previously published data or hypotheses.

- Controversial: findings of the article either challenge the established dogma in a given field, or require further validation before they can be considered irrefutable.

- Good for Teaching: a key article in that field and/or a particularly well written article that provides a good overview of a topic or is an excellent example of which students should be aware.

- Interesting Hypothesis: proposes a novel model or hypothesis that the recommending Faculty Member found worthy of comment.

- New Finding: presents original data, models or hypotheses.

- Novel Drug Target: the article suggests a specific, new therapeutic target for drug discovery (rather than a new drug).

- Refutation: the findings of the article disprove previously published data or hypotheses (where a specific finding is being refuted, a reference citation is normally required).

- Technical Advance: introduces a new practical/theoretical technique, or novel use or combination of an existing technique or techniques.

As a rule, the classification of a paper is not linked to its recommendation by a Faculty member. For example, a "good" paper that introduces a new technique is not less practically relevant than an "exceptional" paper that has the "technical advance" tag. The tags are intended to be an additional filter/classification rather than part of the rating. The tags are very useful because they are only possible to assign if the paper has been read by an expert. One cannot search a literature database (e.g. Web of Science, WoS, Thomson Reuters) for negative results or for clinical practice changing papers using key words. But the human expert-assigned tags enable this in F1000Prime.

The classifications, recommendations and bibliographic information for publications form the fully searchable F1000 database containing more than 100,000 records (end of 2013). Overall, the F1000 database is regarded simply as an aid for scientists to receive pointers to the most relevant papers in their subject area, but also as an important tool for research evaluation purposes. So, for example, Wouters and Costas (2012) write that "the data and indicators provided by F1000 are without doubt rich and valuable, and the tool has a strong potential for research evaluation, being in fact a good complement to alternative metrics for research assessments at different levels (papers, individuals, journals, etc.)" (p. 14).

**2.2   Formation of the dataset, to which bibliometric data and altmetrics are attached**

In January 2014, F1000 provided me with data on all recommendations (and classifications) made and the bibliographic information for the corresponding papers in their system (n=149,227 records). The dataset contains a total of 104,633 different DOIs, among which all are individual papers with very few exceptions. The approximately 30% reduction of the dataset with the identification of unique DOIs can mainly be attributed to the fact that many papers received several recommendations from members and therefore appear multiply in the dataset.

For bibliometric analysis in the current study, citation counts (between publication and the end of 2012) and other bibliometric data (such as WoS subject categories) were sought for every paper in an in-house database of the Max Planck Society (MPG) based on the WoS and administered by the Max Planck Digital Library (MPDL). In order to be able to create a link between the individual papers and the bibliometric data, two procedures were selected in this study: (1) A total of 90,436 papers in the dataset could be matched with one paper in the in-house database using the DOI. (2) With 4,205 papers of the total of 14,197 remaining papers, although no match could be achieved with the DOI, one could be with name of the first author, the journal, the volume and the issue. Thus bibliometric data was available for 94,641 papers of the 104,633 total (91%). This percentage approximately agrees with the value of 93% named by Waltman and Costas (2014), who used a similar procedure to match data from F1000 with the bibliometric data in their own in-house database.

In April 2014, this dataset with 94,641 papers was extended with the addition of altmetric data. These data come from Altmetric (http://www.altmetric.com/) a start-up that focuses on making article level metrics available. Altmetric tracks and analyses the online activity around scholarly literature. A short description of how Altmetric determines, if a tweet, a blog post or a news article mentions a scientific paper, can be found at http://www.altmetric.com/blog/the-donut-factory/. For 65,535 papers (69%) in the dataset, Altmetric could add a range of altmetric data. 850 of these papers could not be included in this study since the corresponding recommendations by the Faculty members are not yet (fully) available (Bornmann, in press). In addition, Altmetric can only reliably provide data for papers published after 2011. For this reason the dataset is reduced in what follows – where altmetric data is statistically evaluated – to papers from the period after 2011.

Since Altmetric could not add altmetric data for all the papers, but only for 69%, the question arises how the remaining papers should be treated in the statistical evaluation. One could perhaps argue that these papers should be set to zero counts for all altmetrics.

Apparently, not even a mention is available for these papers in any of the social media platforms. On the other hand, the dataset from Altmetric should have at least a mention for all papers under F1000 since the data to which Altmetric attaches altmetric data originate from F1000, and F1000 recommendations are also evaluated by Altmetric. But since this is not the case in the current dataset, the 31% of the papers for which Altmetric was not able to supply any altmetric data were recorded as missing and excluded from the statistical analysis. As the following analyses shows, only a few papers published after 2011 are affected by this problem – that is those papers which were used in the current study for the analysis of alternative metrics.

**Table** 1
Altmetric data made available by Altmetric for papers published after 2011 (n=13,678)

| Number of … | Mean | Minimum | Maximum | Percent of papers with 0 counts |
|---|---|---|---|---|
| unique blogs with post mentioning paper | 0.46 | 0 | 111 | 84 |
| F1000Prime reviews (see section 2.1) | 1 | 0 | 2 | 0 |
| Facebook users or pages mentioning paper | 1.3 | 0 | 433 | 69 |
| number of Google+ users mentioning paper | 0.24 | 0 | 63 | 92 |
| LinkedIn group forums mentioning paper | 0 | 0 | 2 | 100 |
| news outlets | 0.64 | 0 | 128 | 87 |
| peer review sites (Publons, PubPeer) | 0.15 | 0 | 44 | 100 |
| Pinterest users mentioning paper | 0.01 | 0 | 5 | 99 |
| Q&A threads on Stack Exchange, Math Overflow etc. | 0 | 0 | 1 | 100 |
| posts on Reddit (not comments on those posts) | 0.46 | 0 | 19 | 97 |
| unique tweeters mentioning paper | 11.83 | 0 | 2161 | 29 |
| unique YouTube users with a video mentioning the paper | 0.14 | 0 | 8 | 99 |
| Total altmetric counts (sum of all mentions) | 15.52 | 1 | 2745 | 0 |

Table 1 displays the arithmetic mean for the counts, as well as the minimum and maximum for all papers published after 2011 for which Altmetric provides data. In addition, the fraction of papers with 0 counts is also provided. As the results show, the counts in the altmetrics are generally low. For example, the papers in the dataset have an average of 0.46 blogs mentioning a paper with a minimum of 0 and a maximum of 111. An important reason for the generally low averages in the altmetrics is the large fraction with 0 counts: For almost all altmetrics in the table, (significantly) more than two thirds of the papers show zero counts. Since the total altmetric counts and the unique tweeters mentioning papers (Twitter counts) are the only altmetrics with significantly higher average counts and significantly lower share of 0 counts than with the other altmetrics, these are the only ones included in the following statistical analysis.

On the one hand, the analysis of Twitter counts in this study has the further advantage that the data evaluated originates only from one service (which represents the standard in the area of microblogging). This facilitates the collection of data for Altmetric and ensures the reliability of the counts. Blogs, for example, do not have this advantage: "While most other Web 2.0 applications are closely identified with a few 'name-brand' services (for instance, Twitter for microblogging and delicious for social bookmarking), blogging is not" (Priem & Hemminger, 2010). Blogs are distributed over the whole Web, and there is no standard service aggregating these blogs. On the other hand, Twitter is particularly in use by people who operate outside the area of science: Although Twitter is one of the most often used social media platforms, it is generally assumed that only few scientists actually tweet (Darling, et al., 2013; Mahrt, Weller, & Peters, 2012).

### 2.3 Statistical procedure and software used

The statistical software package Stata 13.1 (http://www.stata.com/) is used for this study; in particular, the Stata commands nbreg, margins, and coefplot are used.

A series of regression models has been estimated. The outcome variables (number of citations, number of tweeds, number of total altmetric counts) in the models are count variables. They indicate "how many times something has happened" (Long & Freese, 2006, p. 350). The Poisson distribution is often used to model information on counts. However, this distribution rarely fits in the statistical analysis of bibliometric and altmetric data, due to overdispersion. "That is, the [Poisson] model underfits the amount of dispersion in the outcome" (Long & Freese, 2006, p. 372). Since the standard model to account for overdispersion is the negative binomial (Hausman, Hall, & Griliches, 1984), negative binomial regression models are calculated in the present study (Hilbe, 2007).

The violation of the assumption of independent observations by including several different items of information about the same paper (such as several F1000 recommendation scores or several subject categories associated with a paper) is considered by using the cluster option in Stata (StataCorp., 2013). This option specifies that the information items are independent across papers but are not necessarily independent within the same paper (Hosmer & Lemeshow, 2000, section 8.3).

The publication years of the papers were included in the models predicting different counts (e.g. citations) as exposure time (Long & Freese, 2006, pp. 370-372). The exposure option provided in Stata takes into account the time that a paper is available for citations or other mentions (e.g. in Twitter).

In this study, adjusted predictions are used to make the results easy to understand and interpret. Such predictions are referred to as margins, predictive margins, or adjusted predictions (Bornmann & Williams, 2013; Williams, 2012; Williams & Bornmann, in preparation). The predictions allow a determination of the meaning of the empirical results which goes beyond the statistical significance test. Whereas the regression models illustrate which effects are statistically significant and what the direction of the effects is, adjusted predictions can provide us a practical feel for the substantive significance of the findings.

# 3 Results

## 3.1 The distribution and selection of the tags in the dataset

Table 2 shows the distribution of the tags over the records in the dataset (in which papers appear more than once) and total tag mentions ("total" line). It is very clear that the tags are applied very differently: Whereas, for example, "new finding" makes up about half of the tag mentions, for "review" it is only about 2%. In order to be able to make a reliable statement about the validity of the altmetrics, the following statistical analysis does not include all tags, but only those with more than 5% of mentions or allocated to more than 10% of records.

**Table** 2
Tags, allocated by Faculty members (n=17,805 records, n=25,557 tag mentions). This assessment applies only to papers with a publication year later than 2011, since only these papers are included in the statistical analysis of the altmetrics.

| Tag | Absolute numbers | Percent of tag mentions | Percent of records |
|---|---|---|---|
| New finding | 11,813 | 46.22 | 66.35 |
| Confirmation | 2856 | 11.18 | 16.04 |
| Interesting hypothesis | 2848 | 11.14 | 16 |
| Good for teaching | 2398 | 9.38 | 13.47 |
| Technical advance | 2147 | 8.4 | 12.06 |
| Controversial | 1219 | 4.77 | 6.85 |
| Novel drug target | 1179 | 4.61 | 6.62 |
| Review | 527 | 2.06 | 2.96 |
| Systematic review | 240 | 0.94 | 1.35 |
| Refutation | 179 | 0.7 | 1.01 |
| Clinical trial (non-RCT) | 77 | 0.3 | 0.43 |
| Negative | 74 | 0.29 | 0.42 |
| Total | 25,557 | 100 | 143.54 |

What expectations are there in the current study in relation to the connection between altmetrics counts or citation counts and the categorization of papers with the five selected tags (which are described in further detail in section 2.1)? In connection with "new finding", "confirmation" and "interesting hypothesis", it is expected that the citation counts for such papers would be higher for those where a Faculty member has used this tag than for those where this did not happen. Since these tags particularly relate to aspects which are relevant in a scientific context, it would not be expected that the altmetric tags show this difference between tagged and untagged papers. In contrast to this, we could expect that papers tagged with "good for teaching" would (also) be interesting for a group of people outside science or research. These are papers which are well written, provide an overview of a topic and are well suited for teaching. Therefore, a higher altmetrics count would be expected for papers with this tag than for papers without it. The "technical advance" tag is used on papers that present a new technique or tool (whether that's a lab technique/tool or a clinical one) that make an advance on an existing technique. The tag can be used both for research papers and outside, i.e. clinical or fieldwork. Thus, a similar effect of this tag on altmetric or citation counts would be expected in the statistical analysis.

## 3.2 How do the counts differ for differently tagged papers?

In order to ascertain how total altmetric counts, Twitter counts, and citation counts differ with differently tagged papers, three regression models were calculated with the three counts as dependent variables and the tags as independent variables (see Table 3). Each model includes the individual recommendation scores of the Faculty members alongside the tags. This enables us to ascertain the influence of the tags on the different counts, controlling for the effect of the recommendations. Since the recommendations reflect the quality of the papers, the results of the tags are adjusted for the quality of the papers. In other words: the different results for the tags can hardly be traced back to the differing quality of the papers.

**Table** 3
Dependent and independent variables included in the three negative binomial regression models

| Variable | Mean/ Percent | Standard deviation | Minimum | Maximum |
| --- | --- | --- | --- | --- |

| **Altmetrics (papers published after 2011)** | | | | |
|---|---|---|---|---|
| _Dependent variables_ | | | | |
| Total altmetric counts (model 1) | 22.6 | 88.35 | 1 | 2745 |
| Twitter counts (model 2) | 17.61 | 71.86 | 0 | 2161 |
| _Independent variables_ | | | | |
| New finding | 66% | | 0 | 1 |
| Confirmation | 16% | | 0 | 1 |
| Interesting hypothesis | 16% | | 0 | 1 |
| Good for teaching | 14% | | 0 | 1 |
| Technical advance | 12% | | 0 | 1 |
| Recommendation of Faculty member | | | | |
|    Good | 48% | | 0 | 1 |
|    Very good | 42% | | 0 | 1 |
|    Exceptional | 10% | | 0 | 1 |
| Number of recommendations | n=17,805 | | | |
| Number of papers | n=13,678 | | | |
| **Citation counts (papers published before 2011)** | | | | |
| _Dependent variables_ | | | | |
| Citation counts (model 3) | 92.4 | 166.2 | 0 | 3452 |
| _Independent variables_ | | | | |
| New finding | 72% | | 0 | 1 |
| Confirmation | 18% | | 0 | 1 |
| Interesting hypothesis | 21% | | 0 | 1 |
| Good for teaching | 0.1% | | 0 | 1 |
| Technical advance | 16% | | 0 | 1 |
| Recommendation of Faculty member | | | | |
|    Good | 58% | | 0 | 1 |
|    Very good | 35% | | 0 | 1 |
|    Exceptional | 7% | | 0 | 1 |
| Number of recommendations | n=56,604 | | | |
| Number of papers | n=43,329 | | | |

As Table 3 shows, the three models involve papers from different years: The models with altmetrics as dependent variables can only take into account papers published after 2011 (see above). The model with citation counts as dependent variable only involves papers published before 2011. Since the citation window for the papers extends from the publication year to the end of 2012 in this study, the citation window for papers published after 2011 is too narrow to measure the citation impact reliably (Wang, 2013). The inclusion of papers from before 2011 leads, however, to a shortage of records tagged with "good for teaching" (0.1%, n=181) (see Table 3). The "good for teaching" tag is relatively new for F1000Prime; it was introduced only in 2011. Therefore, it cannot be included in the analysis of the citations.

**Table** 4
Results of three negative binomial regression models

|  | Total altmetric counts (model 1) | Twitter counts (model 2) | Citation counts (model 3) |
|---|---|---|---|
| **Tag** | | | |
| New finding | -0.13 | -0.15 | 0.34*** |
|  | (-1.48) | (-1.56) | (15.02) |
| Confirmation | 0.21 | 0.21 | 0.02 |
|  | (1.17) | (1.18) | (0.69) |
| Interesting hypothesis | -0.06 | -0.10 | -0.02 |
|  | (-0.92) | (-1.32) | (-1.05) |
| Good for teaching | 0.28** | 0.32** |  |
|  | (2.74) | (2.94) |  |
| Technical advance | -0.05 | -0.07 | 0.26*** |
|  | (-0.49) | (-0.68) | (8.67) |
| **Recommendation of Faculty member** | | | |
| Good (reference category) | | | |
| Very good | 0.71*** | 0.74*** | 0.49*** |
|  | (13.26) | (13.24) | (28.72) |
| Exceptional | 1.63*** | 1.67*** | 0.95*** |
|  | (16.48) | (16.90) | (24.32) |
| Constant | -5.07*** | -5.34*** | -3.67*** |
|  | (-66.57) | (-65.60) | (-162.09) |

Notes.
t statistics in parentheses
* $p < 0.05$, ** $p < 0.01$, *** $p < 0.001$

The results of the regression models are shown in Table 4. These are the test statistics and p-values, respectively, for the null hypothesis that an individual coefficient is zero, given that the other variables are in the model. The constant is the negative binomial regression estimate when all variables in the model are evaluated at zero (see the annotated Stata output at http://www.ats.ucla.edu/stat/stata/output/stata_nbreg_output.htm). The predicted numbers of count for the different tags and recommendation scores, resulting from the models, are shown in Figure 1, Figure 2, Figure 3, and Figure 4. Since the predicted numbers of counts depend on the models with all independent variables, they are calculated for the different tags under control of the recommendation scores (and adjusted for quality). In all the models in Table 4, a statistically significant result is seen for the recommendation scores of the Faculty members. Since the coefficients have a positive sign, higher total altmetric counts, Twitter counts, and citation counts are to be expected with better scores. Thus the quality of the papers does not only play an important role for the citation impact, but also for the altmetric counts. The relation between the different recommendation scores and the predicted numbers of counts is presented in Figure 1: It is very clear that citation counts in particular separate the differently evaluated papers.

In the two models for the altmetrics (models 1 and 2), the coefficient for "good for teaching" is statistically significant. Correspondingly, Figure 2 and Figure 3 show higher predicted numbers of counts for papers where this tag is set, than for those papers where this was not the case. For example, we can expect a paper with this tag to have around seven Twitter citations more than one without – if the paper is rated as "very good" by Faculty members and has no other tags. These results for "good for teaching" correspond to the expectations (see above) and indicate that altmetric data (and especially tweets) can indicate papers which are of interest outside of science.

Unfortunately, the "good for teaching" tag could not be included in the model for the citation counts (see above). Therefore, there is a lack of results which could be included in a comparison. Model 3 for the citation counts provides two statistically significant results (see Table 4): Citations are particularly to be expected if a paper presents original data, models or hypotheses (tag: "new finding") or introduces a new practical/ theoretical technique (tag: "technical advance"). Whereas the results for "new finding" correspond with the expectations (see above), the results for "technical advance" can clarify the unspecific expectations formulated above: Papers tagged with "technical advance" seem to involve techniques with relevance for research rather than for areas outside research. For both tags, Figure 4 shows a clear citation impact advantage for papers with this tag than for those without.

The results in Figure 4 also show that confirmatory results (tag: "confirmation") and interesting hypotheses (tag: "hypothesis") can hardly be associated with higher or lower citation counts (against the expectation).

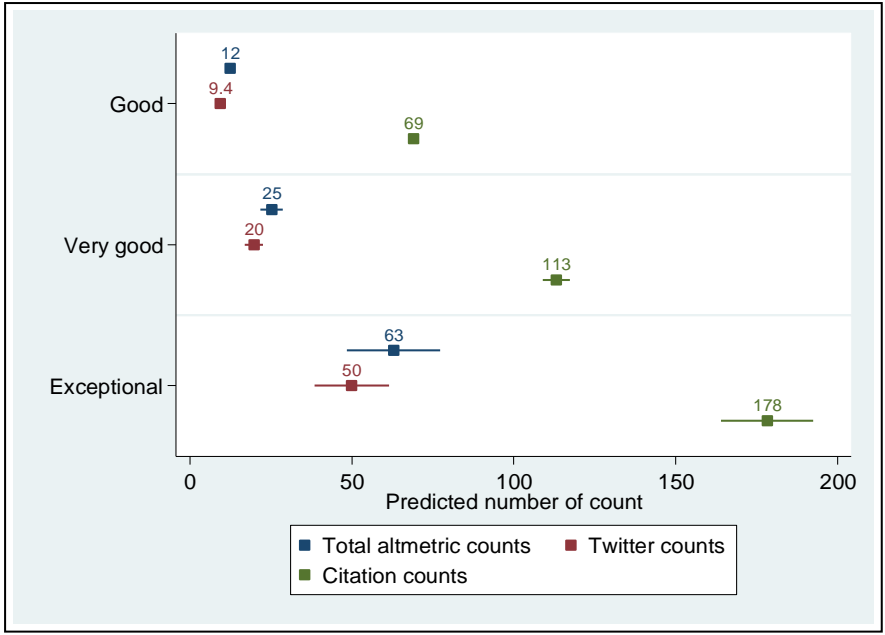

Figure 1. Predicted numbers of total altmetric counts, Twitter counts und citation counts with 95% confidence intervals for three <u>individual recommendation scores.</u> Whereas papers published before 2011 are included in the evaluation of the citation counts, the evaluation of the altmetric data involves papers after 2011.

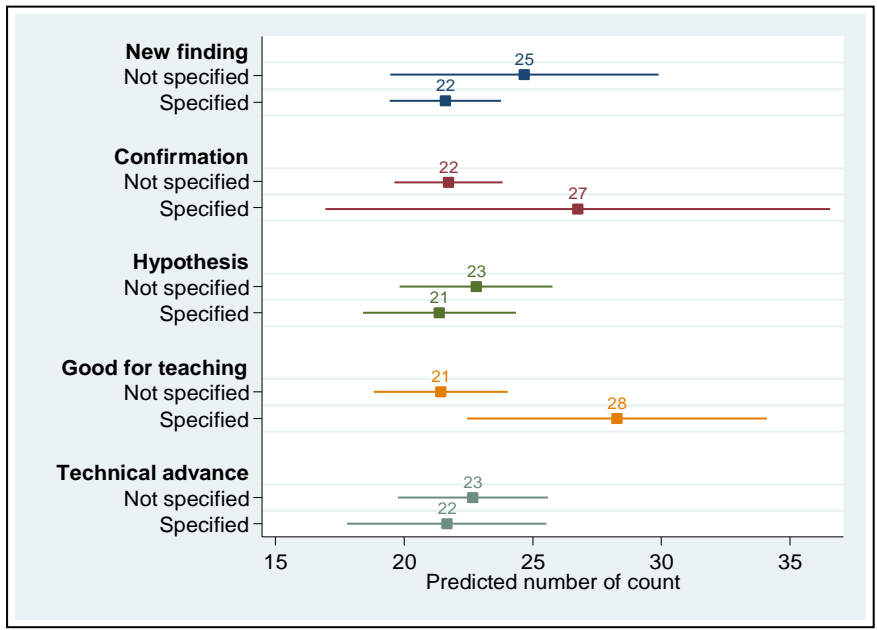

Figure 2. Predicted numbers of total altmetric counts with 95% confidence intervals for papers tagged differently (papers published after 2011)

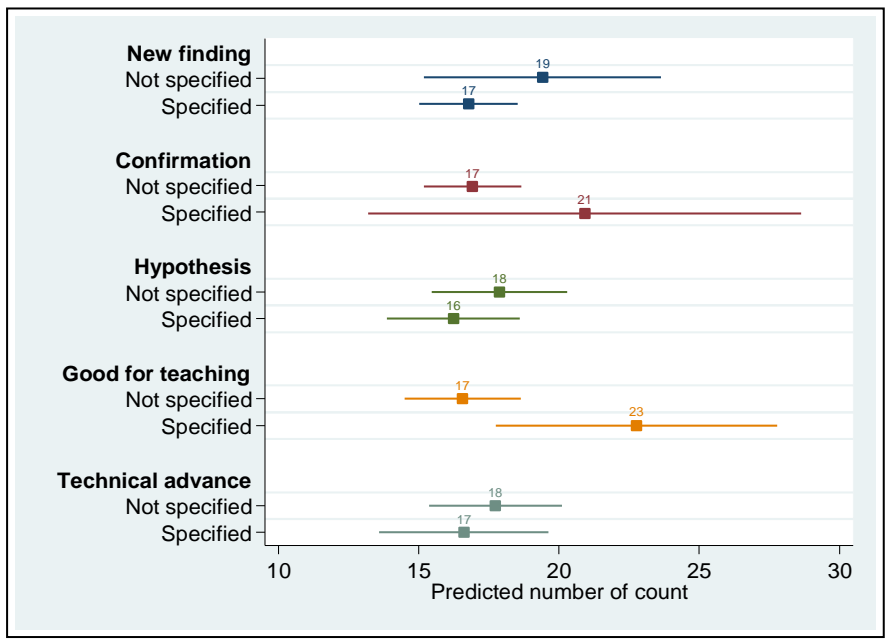

Figure 3. Predicted numbers of Twitter counts with 95% confidence intervals for papers tagged differently (papers published after 2011)

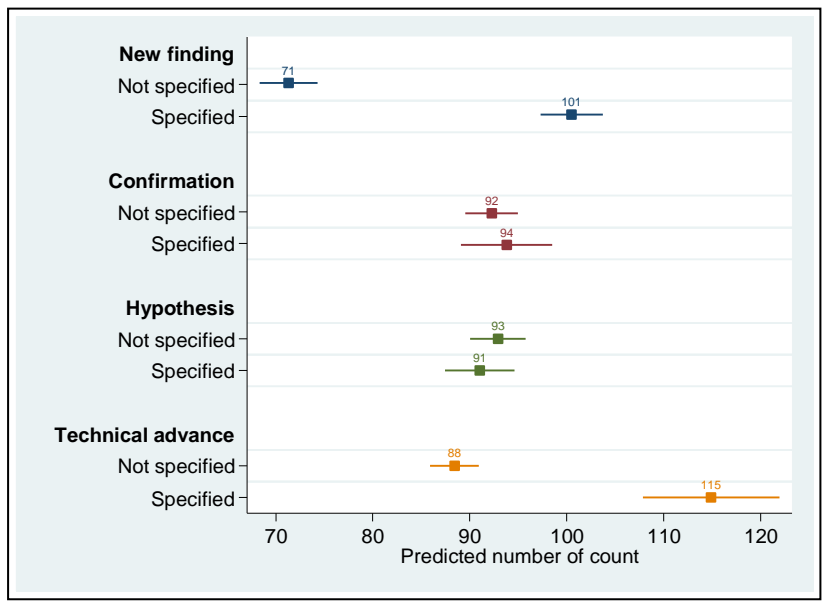

Figure 4. Predicted numbers of citation counts with 95% confidence intervals for papers tagged differently (papers published after 2011)

## 3.3 Is normalization of impact necessary for altmetric counts, as with citation counts?

If altmetric data is to be used for the measurement of societal impact in the evaluation of research, the question arises of its normalization (Torres-Salinas, Cabezas-Clavijo, & Jimenez-Contreras, 2013). With citation counts, there is a consensus in the bibliometric community that the impact of papers should be normalized in relation to the subject category (the field) and the publication year (the time) (Bornmann, Leydesdorff, & Wang, 2013). Is this also necessary for the Twitter counts and total altmetric counts investigated here? The following statistical analysis will focus on the question of taking into account the subject categories, since, for the papers in this study, only the publication year is available and not the publication month or day. Unlike citations which arise only a long time after the appearance of a paper, altmetric data generally appears relatively quickly (Priem, Taraborelli, Groth, & Neylon, 2010; Rodgers & Barbrow, 2013). The temporal aspect in the normalization of altmetric data can therefore only be clarified with data on the month or day level.

Other empirical studies have already indicated subject area differences with altmetric data. Thus, for example Loach (2014) shows from Twitter counts in the Altmetric database "that medical articles receive a disproportionate amount of online attention. In fact, 60% of tracked tweets from the last week pointed to articles from journals publishing Medical and Health Science research. Interestingly, 63% of these were directed to articles from journals tagged as relating to Clinical Medicine or Public Health specifically." An important disadvantage of the studies which have so far appeared on subject area difference is that the quality of the papers is not controlled in the analyses of the subject area differences. Therefore, it is not known whether the differences between the subject areas depend on aspects specific to the subject or quality differences between the papers. Thus, medical papers could receive more online attention just because these papers are generally of a higher quality than papers from other subject categories. The quality of the papers should therefore be controlled in the analysis of subject area differences.

In the current study, WoS subject categories are used to determine subject area differences in the counts. Most bibliometric studies use these categories, which, however, are not applied on the level of individual papers, but on the level of journals: A set of journals is combined in a subject category by Thomson Reuters. Table 5 shows the distribution of the papers over the subject categories published after 2011 in the dataset. Since the evaluation of the altmetric data could only include papers after 2011, the table refers to this part of the data. As the table shows, around 14% of the category classifications relate to "multidisciplinary sciences" – that corresponds to around 20% of the papers. This journal set includes the two multi-disciplinary journals *Nature* and *Science*. Around 13% of the papers in the dataset were published in a journal belonging to the category "cell biology".

**Table** 5
Distribution of the papers over subject categories published after 2011 (n=17,805). Subject categories are only listed if they appear more than 100 times in the dataset.

| Subject category | Number of instances | In percent of the category instances | In percent of papers |
|---|---|---|---|
| Multidisciplinary Sciences | 3,502 | 13.97 | 19.67 |
| Cell Biology | 2,316 | 9.24 | 13.01 |
| Biochemistry & Molecular Biology | 2,102 | 8.38 | 11.81 |
| Neurosciences | 1,113 | 4.44 | 6.25 |
| Medicine, General & Internal | 966 | 3.85 | 5.43 |
| Immunology | 951 | 3.79 | 5.34 |
| Oncology | 674 | 2.69 | 3.79 |
| Medicine, Research & Experimental | 652 | 2.6 | 3.66 |
| Urology & Nephrology | 630 | 2.51 | 3.54 |
| Genetics & Heredity | 599 | 2.39 | 3.36 |
| Anesthesiology | 564 | 2.25 | 3.17 |
| Clinical Neurology | 544 | 2.17 | 3.06 |
| Gastroenterology & Hepatology | 537 | 2.14 | 3.02 |
| Surgery | 494 | 1.97 | 2.77 |
| Microbiology | 482 | 1.92 | 2.71 |
| Cardiac & Cardiovascular System | 471 | 1.88 | 2.65 |
| Endocrinology & Metabolism | 402 | 1.6 | 2.26 |
| Critical Care Medicine | 361 | 1.44 | 2.03 |
| Respiratory System | 356 | 1.42 | 2 |
| Peripheral Vascular Diseases | 318 | 1.27 | 1.79 |
| Obstetrics & Gynecology | 316 | 1.26 | 1.77 |
| Infectious Diseases | 311 | 1.24 | 1.75 |
| Dermatology | 310 | 1.24 | 1.74 |
| Hematology | 306 | 1.22 | 1.72 |
| Developmental Biology | 296 | 1.18 | 1.66 |
| Ophthalmology | 294 | 1.17 | 1.65 |
| Psychiatry | 292 | 1.16 | 1.64 |
| Pediatrics | 289 | 1.15 | 1.62 |
| Pharmacology & Pharmacy | 279 | 1.11 | 1.57 |
| Virology | 254 | 1.01 | 1.43 |
| Ecology | 227 | 0.91 | 1.27 |
| Chemistry, Multidisciplinary | 224 | 0.89 | 1.26 |
| Plant Sciences | 204 | 0.81 | 1.15 |
| Otorhinolaryngology | 191 | 0.76 | 1.07 |
| Biophysics | 188 | 0.75 | 1.06 |
| Reproductive Biology | 188 | 0.75 | 1.06 |
| Parasitology | 180 | 0.72 | 1.01 |
| Rheumatology | 180 | 0.72 | 1.01 |
| Biology | 177 | 0.71 | 0.99 |
| Physiology | 177 | 0.71 | 0.99 |
| Evolutionary Biology | 173 | 0.69 | 0.97 |
| Biotechnology & Applied Microbiology | 164 | 0.65 | 0.92 |
| Allergy | 148 | 0.59 | 0.83 |
| Biochemical Research Methods | 142 | 0.57 | 0.80 |
| Public, Environmental & Occupational Health | 113 | 0.45 | 0.63 |

The subject categories in Table 5 are included as independent variables in two negative binomial regression models, where one includes the total altmetrics counts and the other the Twitter counts as dependent variable. With the help of this model, the predicted numbers of counts could be determined for the individual subject categories, where the quality of the papers is controlled for by the individual recommendation scores (which are included as mean scores per paper in the model alongside subject categories). The model also takes into account that the papers appeared in different publication years and have different numbers of subject categories. The results of the regression models will not be presented in table form in what follows, since the tables are very extensive given the large number of different subject categories. But the predicted numbers of counts with 95% confidence intervals for the individual subject categories are presented as the results of the models (see Figure 5 and Figure 6).

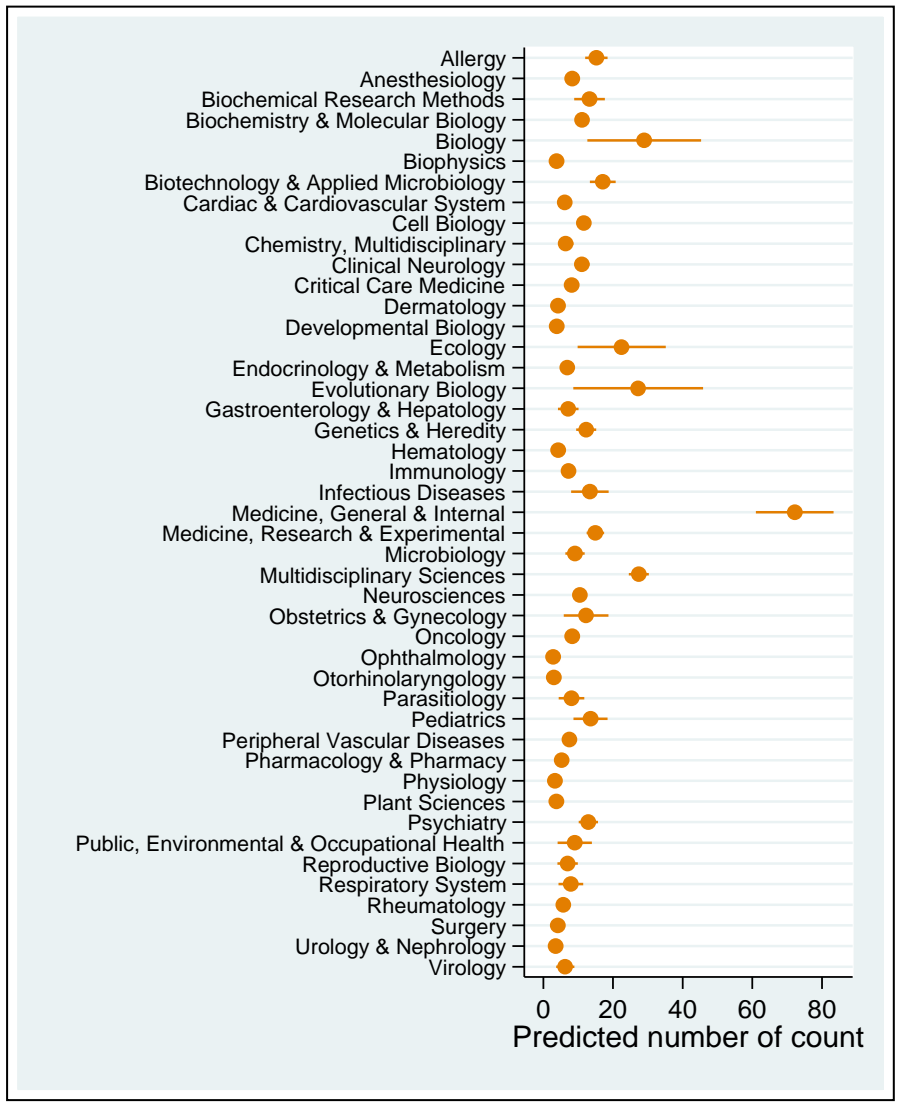

Figure 5. Predicted numbers of total altmetric counts with 95% confidence intervals. The graphic is based on 13,278 papers published after 2011 with 18,254 subject category instances (only subject categories with more than 100 instances). The results arise from a negative binomial regression model, in which the quality of the papers is controlled by the individual recommendation scores of the Faculty members.

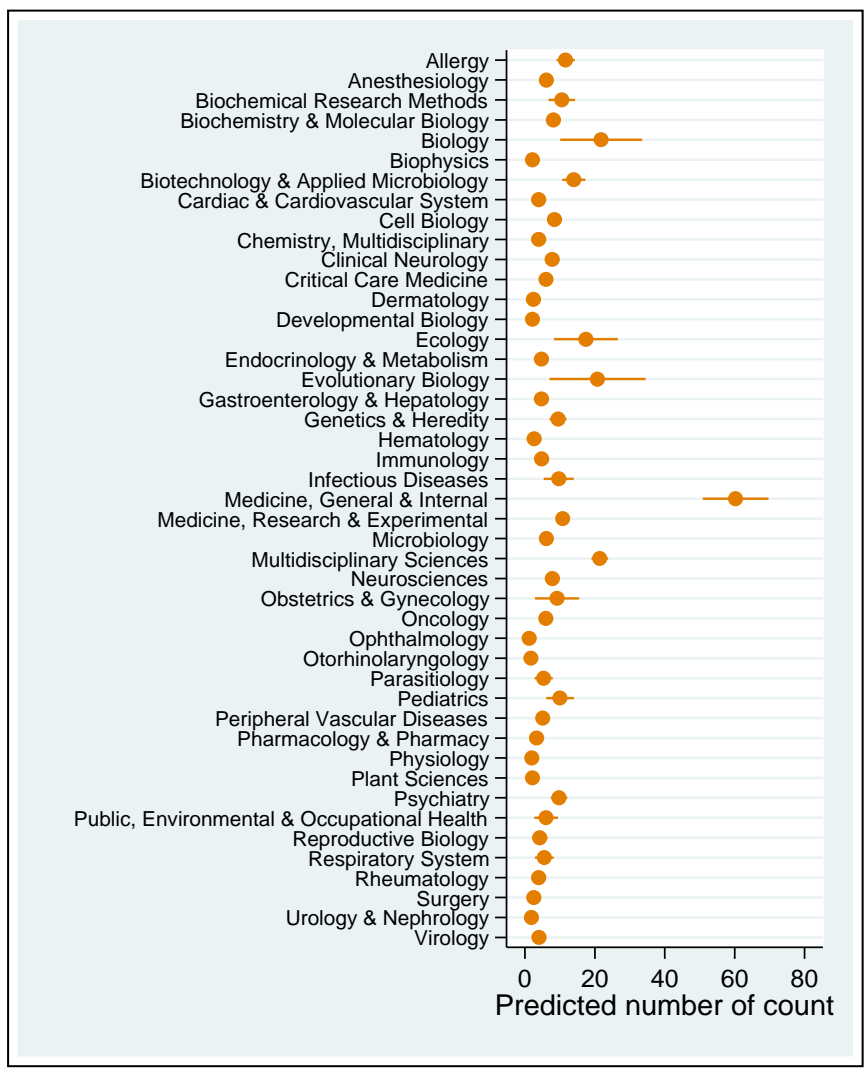

Figure 6. Predicted numbers of Twitter counts with 95% confidence intervals. The graphic is based on 13,278 papers published after 2011 with 18,254 subject category instances (only subject categories with more than 100 instances). The results arise from a negative binomial regression model, where the quality of the papers is controlled for with the individual recommendation scores of the Faculty members.

In order to determine whether the predicted numbers of counts for the subject categories with the altmetric data follows a similar (or different) pattern to that with the citation counts, Figure 7 shows the predicted numbers of citation counts with 95% confidence intervals. As with the evaluations described in section 3.2, these predicted numbers arise from a negative binomial regression model based on papers from the time period before (and not after) 2011. Even if the analysis underlying Figure 7 only took into account subject categories with more than 100 instances in the dataset (similarly to Figure 5 and Figure 6), the subject categories in Figure 7 do not coincide with those shown in Figure 5 and Figure 6. The reason for the discrepancies lies in the different publication years involved.

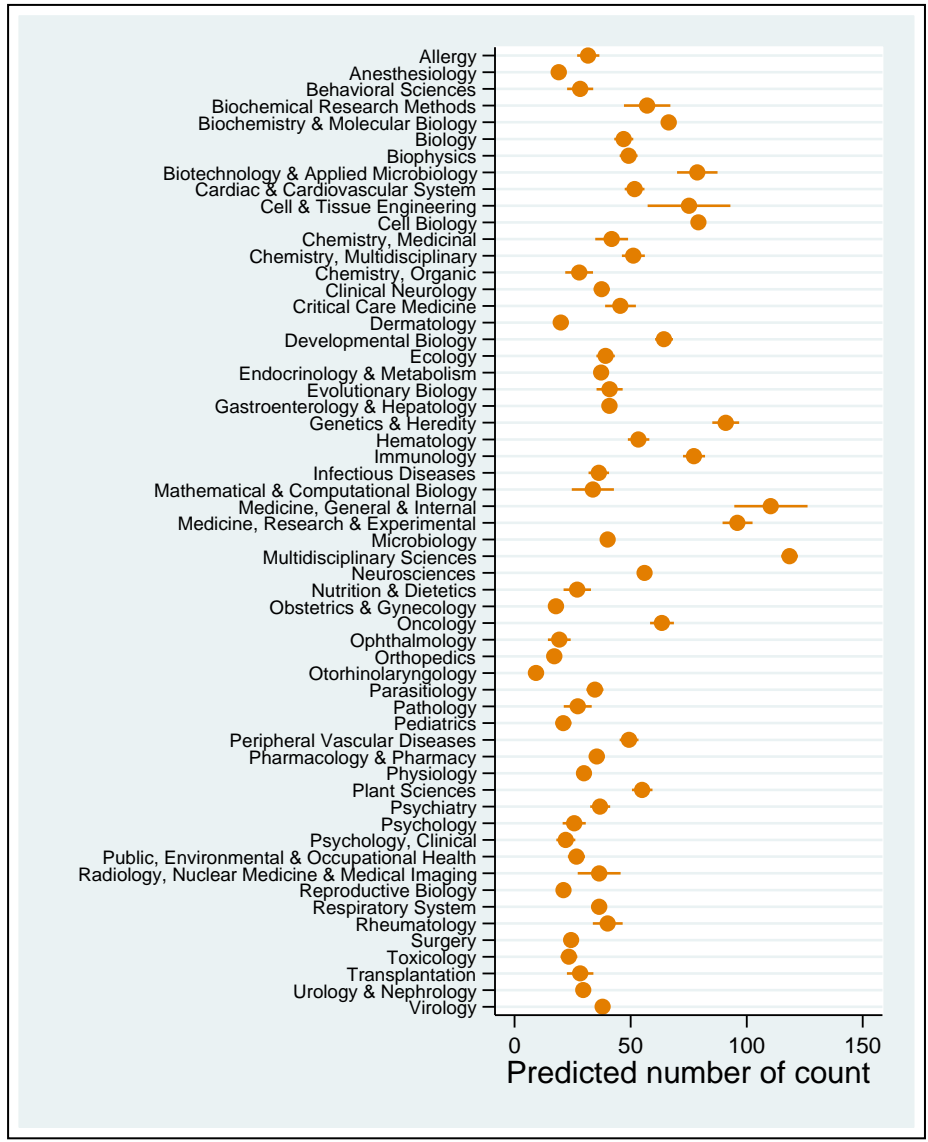

Figure 7. Predicted numbers of citation counts with 95% confidence intervals. The graphic is based on 42,858 papers published before 2011 with 60,468 subject category instances (only subject categories with more than 100 instances). The results arise from a negative binomial regression model, where the quality of the papers is controlled for with the individual recommendation scores of the Faculty members.

As the results in the three figures show, the predicted numbers of counts for the altmetric data on the one hand is very different from that for the bibliometric data, on the other. With the altmetric data (total altmetric counts and Twitter counts), the predicted number of counts is relatively low for almost all subject categories. Only for "biology", "ecology", "evolutionary biology", "multidisciplinary sciences" and especially for "medicine, general & internal" are they higher. Particularly in the journals of the subject category "medicine, general & internal" an especially large number of contributions seem to be published which are not only of scientific interest.

The predicted numbers of citation counts shown in Figure 7, shows a different pattern from the predicted numbers of altmetrics data. In Figure 7 there are quite a few subject categories which stand out with relatively high counts (and many categories with hardly any), but the individual subject categories are distributed over a large bandwidth of different predicted numbers of counts. This difference between the altmetric data and citations in the distribution over the predicted numbers of counts is visualized in Figure 8. Box plots are used to represent the distribution of the counts which are visualized in Figure 5, Figure 6 and Figure 7. In Figure 8 it can clearly be seen that the predicted numbers of citation counts are distributed over a greater area than the predicted numbers of total altmetric counts and Twitter counts. Correspondingly, the citation counts show a significantly greater standard deviation than the altmetric data (see Figure 8).

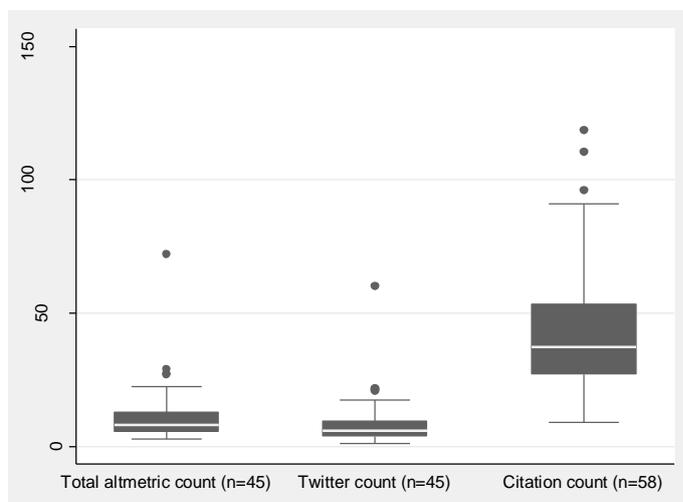

Figure 8. Distribution of the predicted number of total altmetrics counts, Twitter counts and citation counts. Whereas the standard deviations for the total altmetrics counts and Twitter counts are std=11.3 and std=9.5, for the citation counts it is std=23.5.

The results for the differences in the distribution of the predicted numbers of counts between the altmetric and the bibliometric data indicate that the subject categories have a different meaning in this area. Whereas the evaluation of the bibliometric data indicates different citation practices in the fields (which should be taken into account with a normalization), the evaluation of the altmetric data gives the impression that only papers from a few specific subject areas receive a larger number of mentions. With the altmetric data, it does not therefore appear a matter of different habits in the mentioning of papers between the fields, but of a particularly large (or small) interest among people outside science for papers from a few specific areas (or for the bulk of scientific papers). Therefore, a normalization of the counts on the level of subject categories (journal sets) is not regarded as reasonable.

## 4 Discussion

Can altmetric data be validly used for the measurement of societal impact? The current study has sought to answer this question with a comprehensive dataset from very disparate sources (F1000, Altmetric, and an in-house database based on WoS). In the F1000 peer review system, experts attach particular tags to papers which indicate whether a paper could be of interest for science or rather for other segments of society. In this study, these tags were used in an attempt to analyze the validity of altmetric data. A "good for teaching" tag indicates that a paper could be of interest outside the science. If papers with this tag receive more altmetric counts than those without, this would be an indication of the validity of measuring societal impact with altmetric data. Conversely, papers with tags for specifically scientific aspects (such as "new finding" or "hypothesis") should show no effect on the altmetric counts. For contrast with the altmetric data results, this study analyzed citation counts.

First of all, the results of the regression model in relation to all counts (bibliometric <u>and</u> altmetric) show a correlation with the quality of the papers: With better recommendation scores of the Faculty members, the higher the counts are. For example, for recommendation scores "good", "very good", and "exceptional" the corresponding predicted probabilities of citations are 69, 113, and 178. The effect of the recommendation scores occurs – as expected – more strongly with the citation counts than with the altmetric data, and is in agreement with the results of Bornmann (in press). In the study of Bornmann (in press), it is shown that the recommendations of the Faculty members are correlated with field- und time-normalized citation impact scores. The further results of the regression models show substantial differences between altmetrics counts and citation counts (see Bar-Ilan, 2012). With regard to a possible <u>societal impact measurement</u> with altmetrics, the results of the present study indicate that with altmetric data impact measurement beyond the science seems possible: Papers with the tag "good for teaching" do really achieve higher altmetric counts than papers without this tag – if the quality of the papers is controlled. At the same time, a higher citation count is shown especially by papers with a tag that is specifically scientifically oriented ("new finding"). Although the tag "good for teaching" could not be included in the model with the citation counts (so the contrasting comparison was absent), no (statistically) significant effect was demonstrated for the tag "new finding" in the models with the altmetric data.

The results of this study possibly indicate that papers tailored for a readership outside the area of research or science lead to societal impact. This result is in agreement with the proposal of Bornmann and Marx (2014). To produce societal impact, the authors suggest that scientists write assessment reports summarizing the status of the research on a certain subject and representing knowledge which is available for society to access. An assessment report should be couched in generally understandable terms so that readers who are not familiar with the subject area or the scientific discipline can make sense of it. In the view of Bornmann and Marx (2014), these reports could be seen as part of the secondary literature of science, which has up to now drawn on review journals, monographs, handbooks and textbooks (primary literature is made up of the publications of the original research literature). With the help of these assessment reports it should be possible to reach people from other segments of society (besides science) and to achieve a correspondingly high impact that would then have an effect on the altmetric data.

If altmetric data is to be used for the measurement of societal impact, the question arises of its normalization. Bibliometric data – citations – are normalized for subject area and time. This study has therefore taken a second analytic step involving a possible subject area normalization of altmetric data. In this analysis too, additional citation data was considered to be able to determine common factors and differences in the results. In contrast to the predicted numbers of citation counts, where the subject categories each showed very different clustering, the predicted numbers of altmetric counts (total Altmetric counts and Twitter counts) showed very few subject categories producing high levels of clustering (and the great bulk of the categories low clustering): "biology", "ecology", "evolutionary biology", "multidisciplinary sciences" and especially "medicine, general & internal".

In comparison with the other subject categories (which obtained relatively low counts), these categories are of the sort which appeal to a wider audience public. This wider audience is generally especially interested in topics like ecology and evolution, as well as research results from (internal) medicine (or particular diseases). In addition, there is a special interest in contributions from the best-known scientific journals *Nature*, *Science*, and *Proceedings of the National Academy of Sciences* (PNAS), which publish research from all disciplines. There are obviously – as the results of this study show – particular topics in the biomedical area which are of especially great interest for a wide audience. Since these more or less interesting topics are not completely reflected in Thomson Reuters' journal sets, a normalization of altmetric data (especially Twitter) should not be based on the level of subject categories, but on the level of topics: Thus, for example, Twitter's homepage includes a current list of trending topics as a main feature. "These terms reflect the topics that are being discussed most at that moment on the site's fast-flowing stream of tweets. In order to avoid topics that are popular regularly (e.g., good morning or good night on certain times of the day), Twitter focuses on topics that are being discussed much more than usual, that is, topics that recently experienced an increase of use, therefore trending" (Zubiaga, Spina, Martínez, & Fresno, 2014). The comparison of Twitter citations of papers published on a particular topic would then show a greater or lesser interest in papers on this topic. A normalization of Twitter citations could then be performed on the level of papers on a topic.

## 5      Conclusions

In relation to the measurement of societal impact, the results of this study are promising: Altmetric data (Twitter counts) seem able to indicate papers which produce societal impact. However, it is not clear which kind of impact is measured: Does it measure social, cultural, environmental and/ or economic impact? With evaluating citations in university text books (impact on education), patents (impact on industry) and clinical guidelines (impact on clinical praxis), there are already some approved instruments available for the reliable societal impact measurements which could be complemented by altmetrics.

In a bid to measure the influence of research on industry, Narin, Hamilton, and Olivastro (1997) studied the frequency with which scientific <u>publications were cited in US patents</u>. They evaluated 400,000 US patents issued between 1987 and 1994. Their results show that the knowledge flow from US science to US industry tripled in these years. Grant (1999) and Lewison and Sullivan (2008) pursued a similar objective to Narin, et al. (1997) with their evaluation of <u>clinical guidelines</u>: how does knowledge flow from clinical research to clinical practice? The pilot study by Grant (1999) examined three guidelines and was able to ascertain that they contained citations of a total of 284 publications (which can be categorised by author, research institution, country, etc.). For Grant (1999), the study results demonstrate the usefulness of his approach to tracing the flow of knowledge from research funding into clinical practice.

As most of the former empirical studies on altmetrics have pointed out, we need further studies (including a broad range of altmetrics) dealing with the question of the specific impacts of altmetrics. For this, datasets are required which contain information about the importance of individual publications outside the area of science. This information should be produced by experts (and thus be reliable and valid). Unfortunately, the F1000 dataset does not contain this information. It would be particularly interesting to have information on the importance of publications for very specific segments of society (such as the economy, politics or culture). With this data, one could determine which specific altmetric impact one could measure in which segment of society.

# Acknowledgements

I would like to thank Adie Chan, Ros Dignon, and Antonia Desmond from F1000 for providing me with the F1000Prime data set and for providing feedback on this study. Furthermore, I would like to thank Euan Adie from Altmetric for providing me with altmetric data and for comments on the study. The bibliometric data used in this paper are from an in-house database developed and maintained by the Max Planck Digital Library (MPDL, Munich) and derived from the Science Citation Index Expanded (SCI-E), Social Sciences Citation Index (SSCI), Arts and Humanities Citation Index (AHCI) prepared by Thomson Reuters (Scientific) Inc. (TR®), Philadelphia, Pennsylvania, USA: ©Copyright Thomson Reuters (Scientific) 2014.